 \documentclass[final,3p,times]{elsarticle}
\usepackage{amssymb}
\usepackage{amsmath}

\usepackage{graphicx}

\usepackage{subfig}

\usepackage{axodraw4j}

\usepackage{booktabs}

\usepackage{color}

\usepackage{braket}

\newcommand{\be}{\begin{equation}}
\newcommand{\ee}{\end{equation}}
\newcommand{\beeq}{\begin{eqnarray}} 
\newcommand{\eeeq}{\end{eqnarray}} 

\newcommand{\uvec}[1]{\underline{#1}}

\begin{document}

\begin{frontmatter}

%% Title, authors and addresses

%% use the tnoteref command within \title for footnotes;
%% use the tnotetext command for the associated footnote;
%% use the fnref command within \author or \address for footnotes;
%% use the fntext command for the associated footnote;
%% use the corref command within \author for corresponding author footnotes;
%% use the cortext command for the associated footnote;
%% use the ead command for the email address,
%% and the form \ead[url] for the home page:
%%
%% \title{Title\tnoteref{label1}}
%% \tnotetext[label1]{}
%% \author{Name\corref{cor1}\fnref{label2}}
%% \ead{email address}
%% \ead[url]{home page}
%% \fntext[label2]{}
%% \cortext[cor1]{}
%% \address{Address\fnref{label3}}
%% \fntext[label3]{}

\title{Recursion relations and scattering amplitudes in  the light-front formalism}

%% use optional labels to link authors explicitly to addresses:
%% \author[label1,label2]{<author name>}
%% \address[label1]{<address>}
%% \address[label2]{<address>}

\author{C.A. Cruz-Santiago}
\address{Physics Department, 104 Davey Lab, The Pennsylvania State University, University Park, PA 16802, USA}
\author{A.M. Sta\'sto}
\address{Physics Department, 104 Davey Lab, The Pennsylvania State University, University Park, PA 16802, USA}
\address{RIKEN  Center, Brookhaven National Laboratory, Upton, NY 11973, USA}
\address{H. Niewodnicza\'nski Institute of Nuclear Physics, Polish Academy of Sciences, ul. Radzikowskiego 152, 31-342 Krak\'ow, Poland}

\begin{abstract}
The fragmentation functions and scattering amplitudes are investigated in the framework of light-front perturbation theory. It is demonstrated that, the factorization property of the fragmentation functions implies the recursion relations for the off-shell  scattering amplitudes which are light-front analogs of the Berends-Giele relations. 
These recursion relations on the light-front can be solved exactly by induction and it is shown that the expressions for the off-shell light-front amplitudes are represented as a linear combinations of the on-shell amplitudes. 
By putting external particles on-shell we recover the scattering amplitudes previously derived in the literature.
\end{abstract}
\end{frontmatter}

%%
%% Start line numbering here if you want
%%
% \linenumbers

%% main text

\section{Introduction}

  Quantization procedure on the  the light-front (or the null-plane)  was first proposed a long time ago by Dirac \cite{Dirac:1949cp} as an alternative approach to the more standard instant-time quantization. One of the interesting features of the light-front quantization is the presence of only three dynamical  Poincar\'e generators which describe the evolution of a system in light-front time, see for example \cite{Leutwyler:1977vy}. Thus, one may hope that the light-front formalism may lead to a simpler solution of problems in relativistic quantum mechanics than other quantization schemes which typically possess  larger number of dynamical operators.  It can be also shown that, there exists a subgroup on the light-front which exhibits algebraic structure   isomorphic to  the Galilean 
symmetry group of non-relativistic quantum mechanics in two dimensions \cite{Susskind:1967rg,Kogut:1972di}.  It has been also argued, see for example \cite{Heinzl:2000ht},  that
the vacuum on the light-front is essentially structureless (the arguments about simplicity of the vacuum   have been provided earlier by analysis of graphs in the infinite momentum frame \cite{Weinberg:1966jm}). This stems from the fact that the lines in the diagrams for amplitudes only have positive $p^+$ momenta which dramatically reduces the number of diagrams which are needed to be considered and eliminates vacuum graphs. To be precise, the vacuum on the light-front is structureless up to zero modes, for which special treatment may be necessary, like discrete quantization which isolates these modes, see for example \cite{Yamawaki:1998cy}. It was also shown that zero modes  contribute to the Higgs VEV in the standard model \cite{Srivastava:2002mw}. Due to the apparent simplicity of vacuum, light-front methods have been also used to study the chiral symmetry breakdown, for a recent nice review see \cite{Beane:2013ksa}.
In any case this property of the light-front vacuum  allows to define unambiguously  the partonic content of hadrons and of hadronic wave functions and  has been used to argue about the presence of  in-hadron quark condensates \cite{Brodsky:2010xf}. The light-front framework has been used   to investigate the hadron dynamics  from AdS/CFT correspondence \cite{Brodsky:2003px} and  in the high energy approximation to compute the  soft gluon component of the heavy onium wave function and to 
obtain a  correspondence with the hard Pomeron in QCD \cite{Mueller:1993rr}.  

In the previous works \cite{ms,CruzSantiago:2013wz}  we have investigated in some detail the gluon wave functions, fragmentation functions and scattering amplitudes within the framework of the light-front perturbation theory (LFPT). The  wave functions differ from fragmentation functions in a way the light-front energy denominators are treated. In the case of the wave functions the energy denominators are kept for the last state ( which is kept off-shell). For the fragmentation functions, the last state is on-shell whereas the first incoming particle is off-shell and the corresponding energy denominator is non-zero. Otherwise there exists close relations between these two objects as they both 
 describe transitions for $1\rightarrow n$ particles.
It has been shown in  \cite{ms,CruzSantiago:2013wz} that one can construct   the recursion relations for each of these objects ( related factorization properties  or cluster decomposition of the light front wave functions were discussed earlier in \cite{Brodsky:1985gs} and the ladder relations between different Fock state components were constructed in \cite{Antonuccio:1997tw} ).
In simpler cases, some of these recursion relations have been solved exactly and the solution for arbitrary number of gluons in the wave function and fragmentation functions has been found. This was done for the case of gluon wave functions and fragmentation functions in the case 
 when the helicities of the outgoing gluons are the same. Interestingly, the compact recursion relations for the gluon wave functions derived   are  related to the vanishing property of the on-shell
helicity amplitudes for these selected configurations of the helicities. It turns out that the property of  vanishing of the amplitudes  for special cases of the helicities,  which was proven \cite{Grisaru:1976vm,Grisaru:1977px} using supersymmetry relations, in the light-front formalism originates from the angular momentum  \cite{Brodsky:2000ii,Harindranath:1998ve,Ma:1998ar} and energy conservation laws. The recursion relations were then generalized to include a different configuration of helicities,
and in the case of the fragmentation functions they are light-front analogs of the   Berends-Giele recursion relations \cite{Berends:1987me}.
 It was also shown that there are general relations between gluon wave functions and scattering amplitudes. More precisely, it was demonstrated that 
 the amplitude $M$ for $2\to n$  can be obtained through analytical continuation from  the  light-front wave functions $\Psi_{n+1}$, which contain, in general, a smaller number of graphs.  
 This property was used to reproduce some lowest order results for the scattering amplitudes which were previously available in the literature, \cite{Parke:1986gb}.  
 
  Despite the fact that the  there has been an enormous progress in the computation of the multi-particle
helicity amplitudes in QCD and many results are now implemented into numerical automated algorithms, (see  for example \cite{Mangano:2002ea,Gleisberg:2008fv,Kleiss:2010hy,Cafarella:2007pc,Dixon:2010ik,Alwall:2011uj} ), it is still  an interesting question to ask whether the obtained results for the scattering amplitudes can be derived in a simpler way in the light-front theory, thus giving a better insight into the structure of the theory.  For example, one of the interesting aspects of using the LFPT is the fact that the variables used to express the helicity amplitudes naturally arise in this framework.
 
In this paper we make significant progress towards answering this question by showing how to derive the maximally helicity violating (MHV) amplitudes in the framework of light-front perturbation theory. We will use the previously constructed recursion relations for the fragmentation functions,  
which are then solved exactly. The solution for the (off-shell) scattering amplitude is proved by mathematical induction. The method, of course, can be used to find the amplitudes for arbitrary configuration of helicities of outgoing particles. The obtained result is more general than the  on-shell amplitude, since it is the solution for an off-shell amplitude with the non-vanishing energy denominator in the first state.  Interestingly, it is expressed as a linear combination of the on-shell amplitudes with different number of external legs, with the first term being proportional to the on-shell amplitude with maximum number of external legs and the subsequent terms containing energy denominator terms. These terms vanish when evaluating the on-shell amplitude, thus reproducing the exact MHV result.

The structure of this paper is the following. In the next section we shall set up a useful notation, and discuss some preliminaries about light-front calculations. We shall recall previously derived  factorized recursion relations for the fragmentation functions, which are the light-front analogs of Berends-Giele recursion relations.
In Sec.~3 recursion relations for the off-shell amplitudes are constructed, which are directly related to the recursion relations for the light-front fragmentation functions. In Sec.~4  we present the solution to the recursion relations  and show that it reduces to the MHV amplitude. Finally in Sec.~5 we state the conclusions.

%%%%%%%%%%%%%%%%%%%%%%%%%%%%%%%%%%%
\section{Light-front perturbation theory and fragmentation functions}

\begin{figure}[ht]
\centerline{\includegraphics[width=0.4\textwidth]{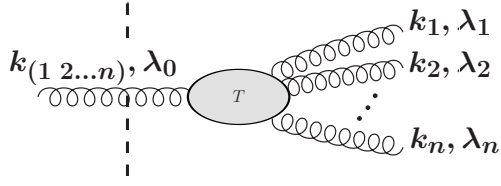}}
\caption{Pictorial representation of the fragmentation amplitude $T_{n}[(1\;2\ldots n)^{\lambda_0} \to 1^{\lambda_1},2^{\lambda_2},\ldots, n^{\lambda_n}]$ for a single off-shell initial gluon. Variables $\lambda_0,\dots,\lambda_n$ denote the polarization of the gluons. The initial gluon $(1\dots n)$ fragments into $n$ final state gluons $1,\dots,n$. The vertical dashed line indicates that for this part of the diagram one needs to take an energy denominator, i.e. the leftmost gluon is in an intermediate state. The other energy denominators which are taken for the intermediate states inside the blob are implicit and are not shown in the picture.}
\label{fig:frag}
\end{figure}

In this section we briefly recall the notation and variables used in the previous works \cite{ms,CruzSantiago:2013wz}
which will be used in the current paper. We will mainly consider transitions from one to $n+1$ final state particles or transitions from $2$ to $n$ final state particles. We shall specialize in this paper to amplitudes with gluons as external particles. 
In this paper we will  use as the building blocks the 'fragmentation functions',
which describe the transition of one particle (ex. a gluon) which is off shell into
$n$ on-shell final state particles. This object, introduced in \cite{ms}, will be denoted by $T_n$ and is depicted in Fig.~\ref{fig:frag}. The off-shell initial gluon is labeled by $(1\dots n)$ and
the final state on-shell gluons are labeled as $1,\dots,n$.
The initial gluon has transverse momentum $\underline{k}_{(1\dots n)}=\sum_{j=1}^n
\underline{k}_j$ and longitudinal momentum fraction $z_{(1\dots n)}=\sum_{j=1}^n z_j$.

%%%%%%%%%%%%%%%%%%%%%%%%%%%%%%%%%%%%
%%%%%%%%%%%%%%%%%%%%%%%%%%%%%%%%%%%%
%%%%%%%%%%%%%%%%%%%%%%%%%%%%%%%%%%%%

The momenta of the last $n$ gluons are labeled $k_1,\dots,k_n$, as in  Fig~\ref{fig:frag}. Each of these momenta can be represented as  $k^{\mu}_i=(z_i P^{+},k_i^-,\underline{k}_i)$, with $z_i$ being the fraction of the initial $P^+$ momentum which is carried by the gluon labeled by $i$ and $\uvec{k}_i$ being the transverse component of the gluon momentum. $P^+$ is the total longitudinal momentum in the initial state, for the example depicted in Fig.~\ref{fig:frag}
it could be the initial state of the total graph to which the subgraph in Fig.~\ref{fig:frag} is attached.
In the 
 LFPT \cite{Weinberg:1966jm,Kogut:1969xa,Bjorken:1970ah,Lepage:1980fj,Brodsky:1997de}  one has  to evaluate the energy denominators for each of the intermediate states for the process.  The energy denominator for say $j$ intermediate gluons  is defined as the difference between the light-front energies of the final and intermediate state in question 
 \be
 {\cal D}_j=\sum_{\rm out} E_{l}-\sum_{i=1}^j E_i  \; .
 \ee
 where 
 \be
 E_{i(l)}\equiv k_{i(l)}^-=\frac{\underline{k}_{i(l)}^2}{k_{i(l)}^+}\; ,
 \ee
  are the light-front energies and the first sum represents a sum over the energies of all final state gluons present in the fragmentation function. Furthermore, one has to sum over all possible vertex orderings.
The  fragmentation function shown in example in Fig.~\ref{fig:frag} would thus
be given schematically by the expression
\be
T_n \sim \sum_{\text{vertex orderings}} g^{n-1} \Pi_{j=1}^{n-1} \frac{V_j}{z_j {\cal D}_j} \;, 
\ee
where $V_j$ are the vertices and $z_j$ and ${\cal D}_j$ are the corresponding fractional momenta and denominators for all the intermediate states. Note the important fact that for the fragmentation function depicted in Fig.~\ref{fig:frag} the first gluon is not really an initial state. As mentioned above, it is understood that the fragmentation function is only a subgraph, attached via this gluon to a bigger graph. Therefore, the leftmost gluon is in fact an intermediate state for which the energy denominator, denoted by the dashed line, has to be taken into account. The rightmost gluons are the final on-shell particles, and the energy denominator is not included there.
Finally, one needs to sum over all the vertex orderings in the light-front time.
The results derived in \cite{ms} and in the following sections are for the color ordered multi-gluon amplitudes. Hence, we focus only on the kinematical parts of the subamplitudes.

The fragmentation function for a special choice of the helicities was evaluated exactly in \cite{ms}. The explicit results for the transition $+\rightarrow +\dots +$ reads
\be
T_n[(12\ldots n)^+ \to 1^+,2^+,\ldots, n^+]=(-ig)^{n-1}\left(\frac{z_{1\ldots n }}{z_1\ldots z_n}\right)^{3/2} \frac{1}{v_{n\;n-1}v_{n-1\;n-2}\ldots v_{21}} \; ,
\label{eq:frag1}
\ee
where the variables $v_{ij}$ were defined as
\be
\uvec{v}_{ij} \; \equiv \; \left( {\uvec{k}_j \over z_j} - {\uvec{k}_i \over z_i}\right)\;, 
\;\;\;\;v_{ij}\equiv \uvec{\epsilon}^{(-)} \cdot \uvec{v}_{ij} \;,
\label{eq:vij}
\ee
and $\uvec{\epsilon}^{(-)}$ will be defined shortly.  It is well known \cite{Susskind:1967rg,Kogut:1969xa,Bjorken:1970ah} that on the light-front the Poincar\'e group can be decomposed onto a subgroup which contains the Galilean-like nonrelativistic dynamics in 2-dimensions. The '$+$' components of the momenta can be interpreted as the 'masses'. In this case
the variable (\ref{eq:vij}) can be interpreted as a relative transverse light-front velocity of the two gluons.  
The same variable is present when evaluating the energy denominators of different intermediate states. The above variable is closely related to 
 the variables 
used in the framework of helicity amplitudes, see \cite{Mangano:1990by}.

For a  given pair of  momenta $k_i$ and $k_j$ we have the result  
\be
\langle ij \rangle = \sqrt{z_i z_j} \; \uvec{\epsilon}^{(-)} 
\cdot \left( {\uvec{k}_i \over z_i} -  {\uvec{k}_j \over z_j} \right)=\sqrt{z_i z_j} \; \uvec{\epsilon}^{(-)}\cdot \uvec{v}_{ij}\, ,
\qquad 
[ij] =  \sqrt{z_i z_j} \; \uvec{\epsilon}^{(+)}\cdot 
\left( {\uvec{k}_i \over z_i} -  {\uvec{k}_j \over z_j} \right)=\sqrt{z_i z_j} \; \uvec{\epsilon}^{(+)}\cdot \uvec{v}_{ij}\, ,
\label{eq:ij}
\ee
where the variables $\langle i  j \rangle$ and  $[ij] $ are defined by
\be
\langle i  j \rangle = \langle i- | j+ \rangle\, ,  \; \; \; [ij] = \langle i+ | j- \rangle \; ,
\label{eq:ijdef1}
\ee
and where  chiral projections of the spinors for massless particles  are defined as
\be
|i \pm \rangle \; = \; \psi_{\pm}(k_i) \; = \; \frac{1}{2}(1\pm\gamma_5)\psi(k_i) \; \; , \;\;\;\;\langle \pm i| \; = \; \overline{\psi_{\pm}(k_i)} \; ,
\label{eq:ijdef2}
\ee
for a given momentum $k_i$. Above, we have also introduced the polarization four-vector
 of the gluon with four-momentum $k$ 
\be
\epsilon^{(\pm)} = \epsilon_{\perp} ^{(\pm)} + 
{2\uvec{\epsilon}^{(\pm)}\cdot \uvec{k} \over \eta\cdot k}\, \eta\; ,
\ee
where
$\epsilon_{\perp} ^{(\pm)} = (0,0,\uvec{\epsilon} ^{(\pm)})$, and
the transverse vector  is defined by 
$\uvec{\epsilon}^{(\pm)} = \mp {1\over \sqrt{2}}(1,\pm i)$.
Vector $\eta$ is related to the choice of the light-cone gauge, $\eta\cdot A = 0$, where 
$\eta^{\;\mu} = (0,2,\uvec{0})$ in the light-front coordinates. 
It is interesting that in the light-front formalism the variables $\langle ij \rangle$ appear naturally in the vertices and in the energy denominators.

\begin{figure}[ht]
\centerline{\includegraphics[width=0.5\textwidth]{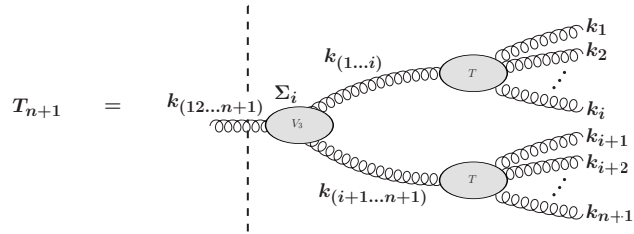}}
\caption{Pictorial representation of the factorization property represented in Eq.~(\ref{eq:fragonestep}), a light-front analog of the Berends-Giele recursion relations \cite{Berends:1987me}. The helicities of the outgoing gluons are chosen to be the same in this particular case. The dashed vertical line indicates the energy denominator $D_{n+1}$.}
\label{fig:fragmaster}
\end{figure}

The fragmentation functions introduced above possess an important property which will be widely utilized in this paper. Namely, it was demonstrated in \cite{ms} that the fragmentation functions  factorize after the summation over all the light-front time orderings. This property can then be used to write down the explicit recursion formula for the fragmentation functions.
That is to say, the fragmentation into $n+1$ gluons which is denoted by $T_{n+1}[(1,2,\ldots, n+1) \to 1,2,\ldots, n+1]$ 
can be represented as the product of two lower fragmentation functions  
$T_i[ (1\ldots i)  \to 1,\ldots,i\,]$ and 
$T_{n+1-i}[ (i+1\ldots n+1)\to i+1,\ldots,n+1]$. Finally, one needs to  sum over the 
splitting combinations. This procedure is schematically expressed in Fig.~\ref{fig:fragmaster} and, to be precise, the expression which reflects the factorization reads
\begin{multline}
T_{n+1}[(12 \ldots n+1) \to 1,2,\ldots,n+1] \; = \; -{2ig\over  D_{n+1}} \; 
\sum _{i=1} ^n  \, \left\{\,
{v^*_{(1\ldots i)(i+1\ldots n+1)} 
\over \sqrt{\xi_{(1\ldots i)(i+1\ldots n+1)}}} \; \right. \\
\left. \times \;\rule{0em}{1.8em}
T_i[(1\ldots i) \to 1,\ldots,i\,] \; T_{n+1-i}[(i+1\ldots n+1) \to i+1,\ldots,n+1]\,
\right\}\, . 
\label{eq:fragonestep}
\end{multline}

%%%%%%%%%%%%%%%%%%%%%%%%%%%%%%%%%%%%
%%%%%%%%%%%%%%%%%%%%%%%%%%%%%%%%%%%%
The energy denominator $D_{n+1}$ in the above equation has been defined as
\be
D_{n+1} \, = \, \frac{\uvec{k}_1^2}{z_1}+ \frac{\uvec{k}_2^2}{z_2}+\ldots +  \frac{\uvec{k}_n^2}{z_n}- \frac{\uvec{k}_{1\ldots n}^2}{z_{1\ldots n}}  \; ,
\ee
i.e. the overall $P^+$ momentum cancels with the normalization of the internal lines (note the difference with respect to the definition of ${\cal D}$ previously).

The following complex representation of the transverse vectors has been introduced
$v_{ij} = \uvec\epsilon^{(-)}\cdot \uvec{v}_{ij}$,  
$v^*_{ij} = \uvec\epsilon^{(+)}\cdot \uvec{v}_{ij}$, and 
along with a useful notation, 
\be
\uvec{v}_{(i_1 i_2 \ldots i_p)(j_1 j_2 \ldots j_q)} = 
{\uvec{k}_{i_1} + \uvec{k}_{i_2} + \ldots + \uvec{k}_{i_p}
\over z_{i_1} + z_{i_2} + \ldots + z_{i_p}} -
{\uvec{k}_{j_1} + \uvec{k}_{j_2} + \ldots + \uvec{k}_{j_q}
\over z_{j_1}+ z_{j_2} + \ldots + z_{j_q}}\; ,
\ee
\be 
\xi_{(i_1 i_2 \ldots i_p)(j_1 j_2 \ldots j_q)} = 
{
(z_{i_1}+z_{i_2}+\ldots+z_{i_p})
(z_{j_1}+z_{j_2}+\ldots+z_{j_q})
\over 
z_{i_1}+ z_{i_2}+ \ldots+ z_{i_p} + z_{j_1} + z_{j_2} + \ldots + z_{j_q}} \;.
\ee
We also introduced notation for the partial sums 
$z_{(1\dots i)}\equiv z_1+z_2+\dots+z_i$ and
$k_{(1\dots i)}\equiv k_1+k_2+\dots+k_i$.

It turns out that the above defined fragmentation functions $T$ are related to the gluonic currents which are building blocks in the Berends-Giele recursion relations \cite{Berends:1987me}.
These recursive relations utilize the (gauge-dependent) current $J^{\mu}$, which is obtained from the scattering amplitudes by putting  one of the particles off-shell. The dual subamplitudes can be obtained by contraction with the polarization vector and setting the gluon back on-shell, \cite{Mangano:1990by}
\be
M(0,1,2,\dots,n) = i P^2\epsilon^{\mu} J_{\mu}(1,2,\dots,n)_{P=-P_0} \; ,
\ee
where we have defined $P=\sum_{i=1}^n p_i$ and in this formula $p_i$ denote the four-vectors for the momenta of the outgoing particles.
In the  light-front perturbation theory the current can also be defined and is related to the fragmentation function as
\be
T_n((12\dots n)\rightarrow 1,2,\dots,n) \equiv \epsilon^{\mu}(12\dots n) \,  J_\mu (1,2,\dots,n) \; ,
\ee
where by $\epsilon^{\mu}(12\dots n)$ we denote the polarization vector of the 
incoming (off-shell) gluon in the fragmentation function.
With such a definition the factorization property for the fragmentation function \eqref{eq:fragonestep} is a light-front analog of the Berends-Giele \cite{Berends:1987me} recursion formula (see also \cite{Kosower:1989xy} for the recurrence relations in the light-cone gauge)
\begin{multline}
J^{\mu}(1,2,\dots,n) = -\frac{i}{P^2}\sum_{i=1}^{n-1} V_3^{\mu \nu \lambda}(p_{1\dots i},p_{i+1\dots n}) J_{\nu}(1,\dots,i) J_{\lambda}(i+1,\dots,n)\\
-\frac{i}{P^2} \sum_{i=j+1}^{n-1} \sum_{j=1}^{n-2} V_4^{\mu \nu \lambda \delta} J_{\nu}(1,\dots,j) J_{\lambda}(j+1,\dots,i) J_{\delta}(i,\dots,n) \; .
\label{eq:bg}
\end{multline}
The simpler form of \eqref{eq:fragonestep} (as compared to \eqref{eq:bg}), which only includes 3-gluon vertex, stems from the fact that it has been written for a particular configuration of helicities, namely for identical helicities of outgoing particles. It is possible to write down a general factorization (recursion) relation for the fragmentation function which will include the 4-gluon vertex as well as the Coulomb term.
We will investigate and use this  more complex case in the next section.
We should also remark that the Berends-Giele relations are  written on the level of individual diagrams, whereas for the derivation of the analogous recursion relations on the light-front
\eqref{eq:fragonestep},  the summation over the different orderings in light-front time  is necessary to decouple the  fragmentation trees.

%%%%%%%%%%%%%%%%%%%%%%%%%%%%%%%%%%%
%%%%%%%%%%%%%%%%%%%%%%%%%%%%%%%%%%%
\section{Recursion relations for off-shell amplitudes on the light-front}

The main goal of this section is to reproduce, within the light-front perturbation theory, the Parke -Taylor \cite{Parke:1986gb} amplitudes by solving the appropriate recurrence relations.
In the following, we will mostly deal with the light-front matrix elements which describe transitions from $1$ to $n+1$ gluons. It has been demonstrated that one can obtain then easily
the amplitudes for $2$ to $n$ transitions from the $1$ to $n+1$ transitions \cite{CruzSantiago:2013wz}, and on the light-front the latter typically involve a smaller number of graphs to evaluate.  The reason for dealing with the matrix elements for $1$ to $n+1$ transitions is that we can directly utilize the factorization property for the fragmentation functions mentioned above.
Thus, following \cite{CruzSantiago:2013wz}, let us introduce the following notation

\begin{align}
{\cal A}_{2\to n}(\{\uvec{k}_0,z_0;\uvec{k}_1,z_1\};\{\uvec{k}_2,z_2;\ldots;\uvec{k}_n,z_n\})&=-{\cal N} M_{1\to n+1}(\{\uvec{k}_0,z_0\};\{\uvec{k}_A,z_A;\uvec{k}_2,z_2;\ldots;\uvec{k}_n,z_n\})\vert_{\uvec{k}_A\to-\uvec{k}_1,\:z_A\to-z_1} \; ,
\label{eq:aandm}
\end{align}

where
 ${\cal A}_{2\to n}$ is the  helicity amplitude for 2 particles going to $n$ particles, for which one needs to set the initial and final state particles as on-shell, and ${\cal N}$ is a normalization factor which will be specified later.
 %%%%%%%%%%%%
 \begin{figure}[h]
\centering
\subfloat[]{\label{fig:1ton+1denom}\includegraphics[width=.5\textwidth]{{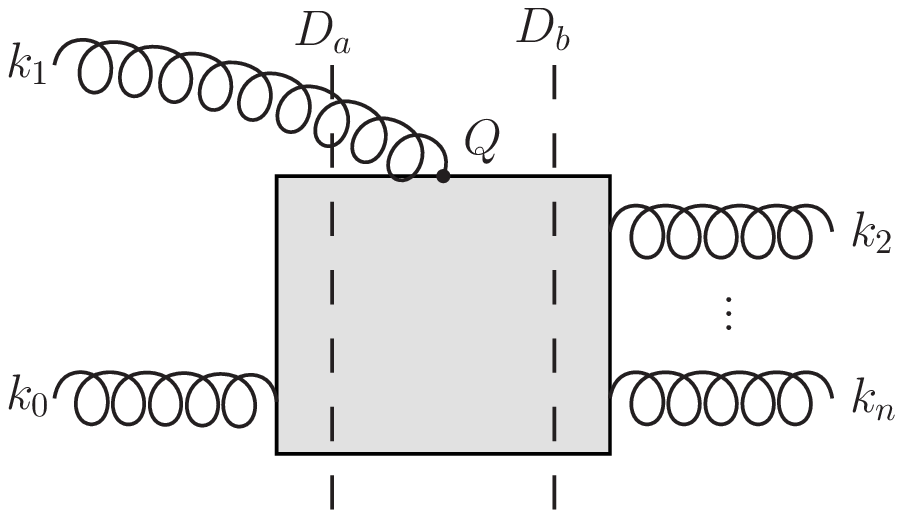}}}
\;\;\;\subfloat[]{\label{fig:2tondenom}\includegraphics[width=.5\textwidth]{{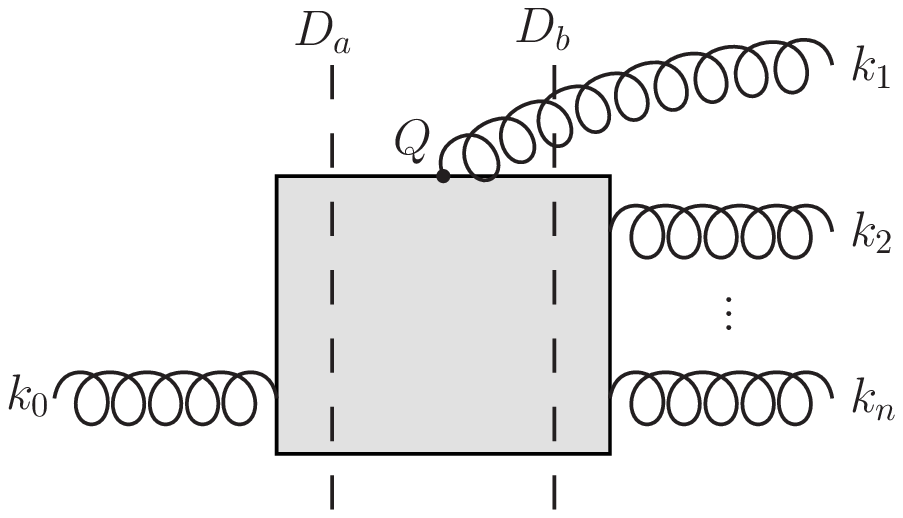}}}
\caption{Schematic representation of the crossing of the momentum of gluon 1 to relate the graphs for $1\to n+1$ with graphs $2\to n$. Point Q denotes the vertex at which gluon 1 attaches to the graph.  $D_a$ and $D_b$ denote the energy denominators for the intermediate states.}
\label{fig:aandm}
\end{figure} 
  %%%%%%%%%%%%

\begin{figure}[h]
\centering
\subfloat[]{\label{fig:recurgrapha}\includegraphics[width=.488\textwidth]{{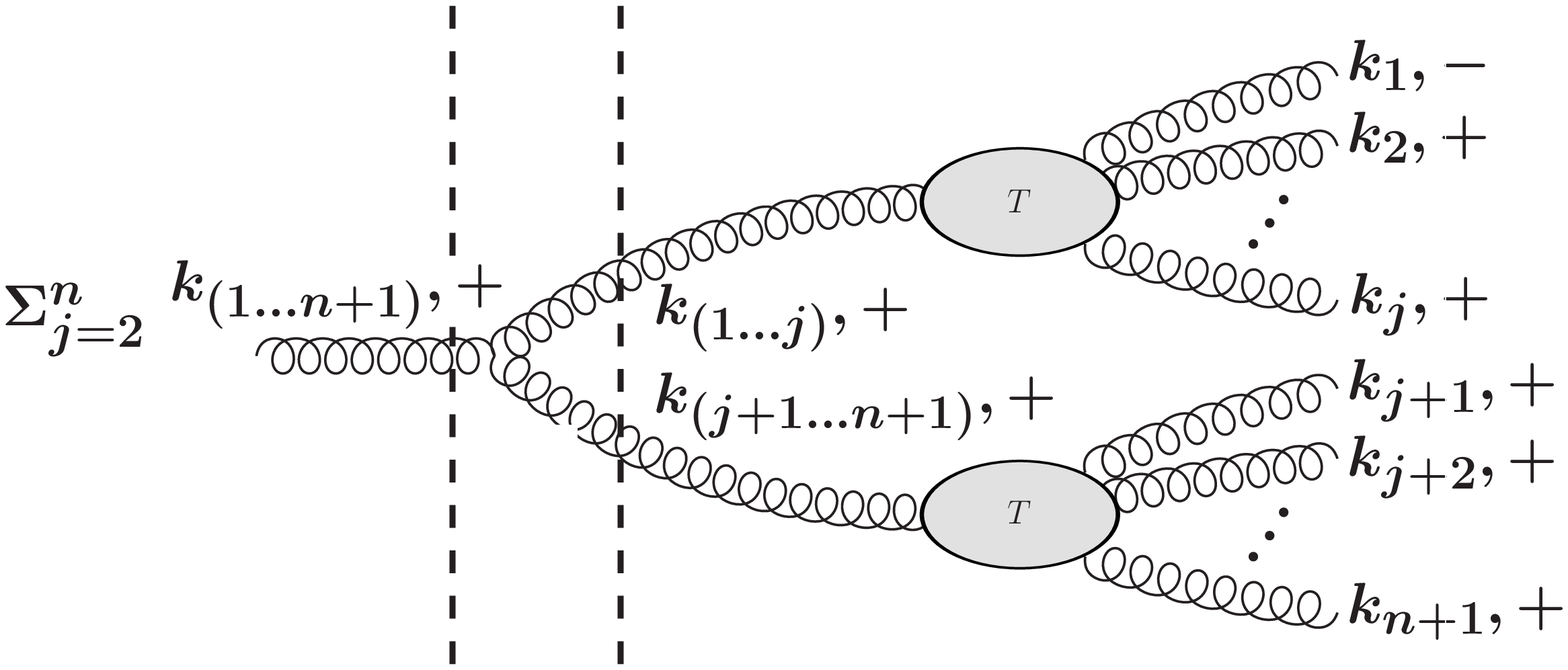}}} \;\;\;
\subfloat[]{\label{fig:recurgraphb}\includegraphics[width=.488\textwidth]{{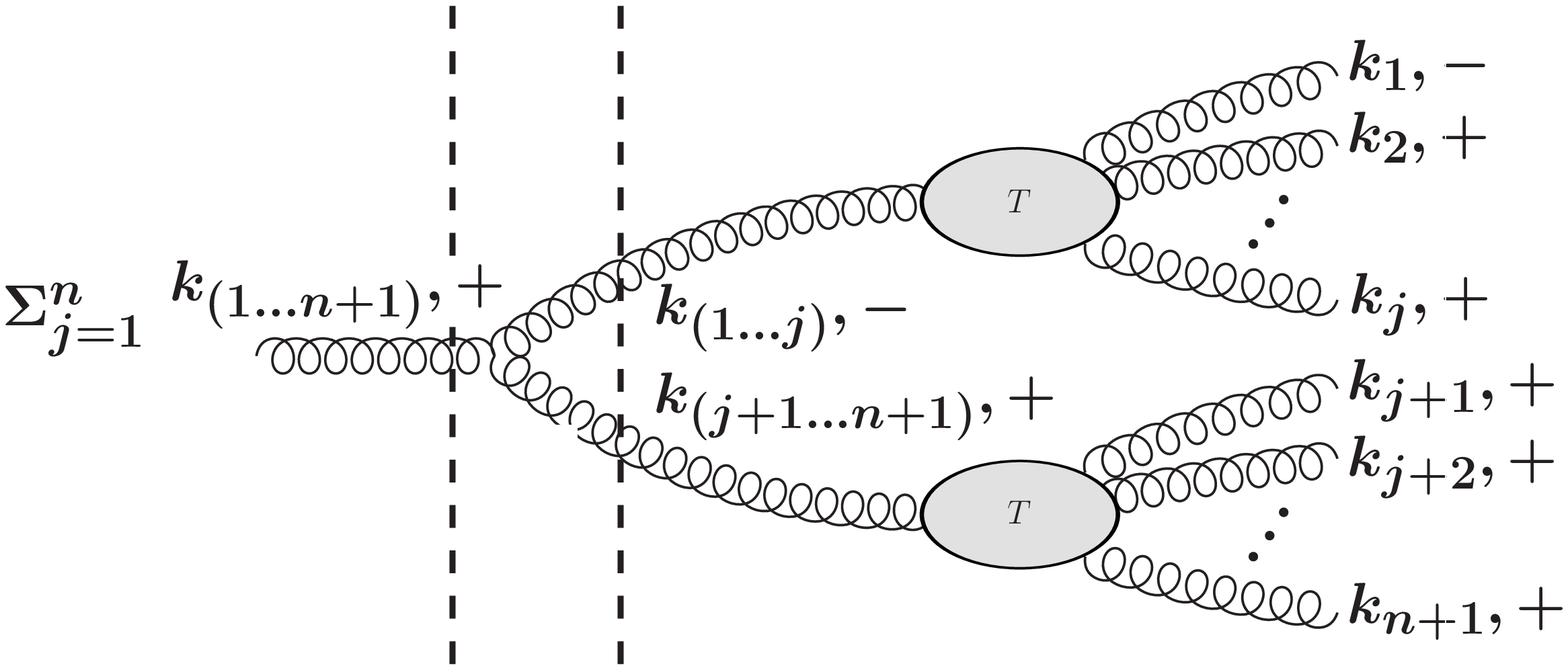}}} \\
\subfloat[]{\label{fig:recurgraphc}\includegraphics[width=.488\textwidth]{{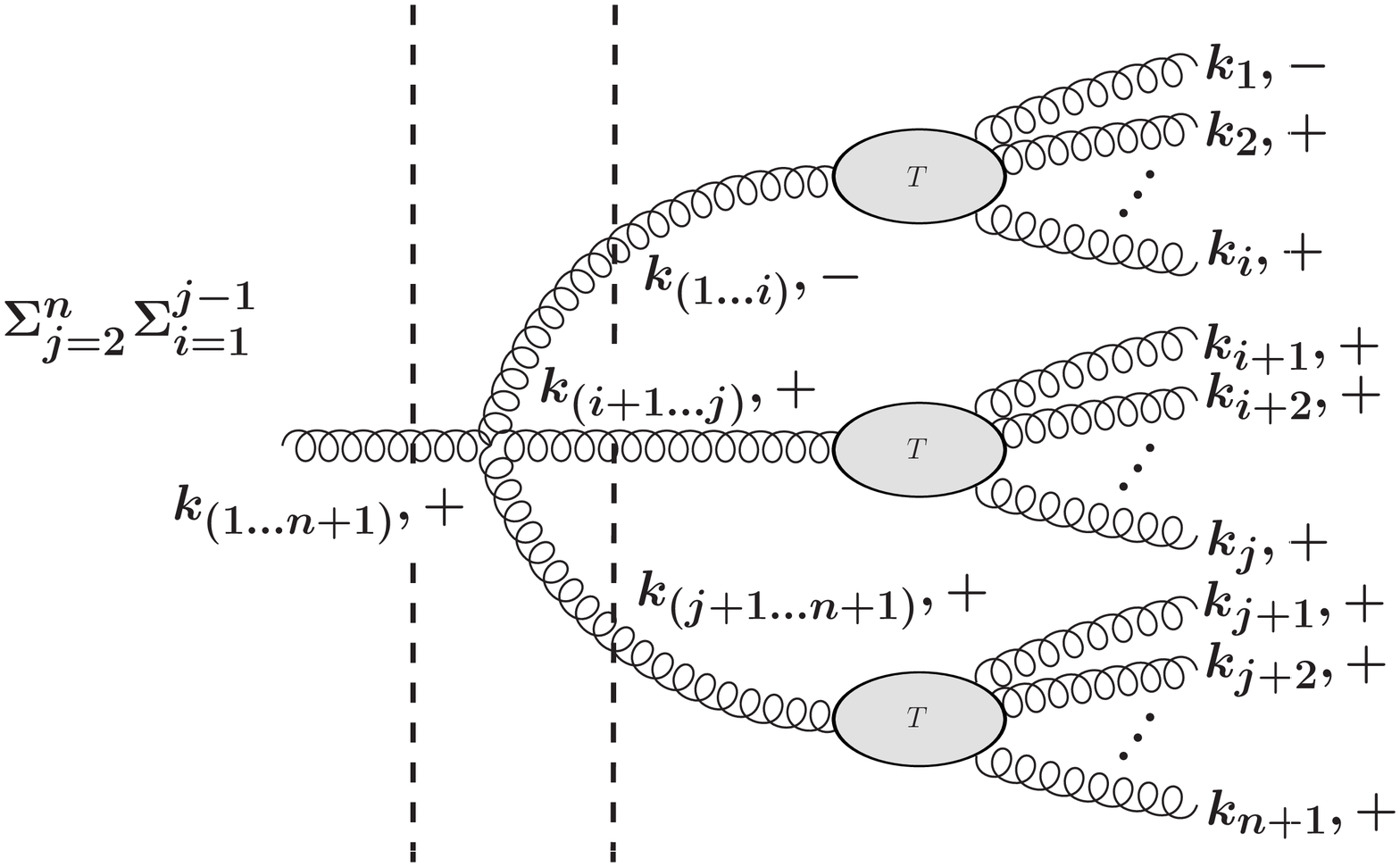}}} \;\;\;
\subfloat[]{\label{fig:recurgraphd}\includegraphics[width=.488\textwidth]{{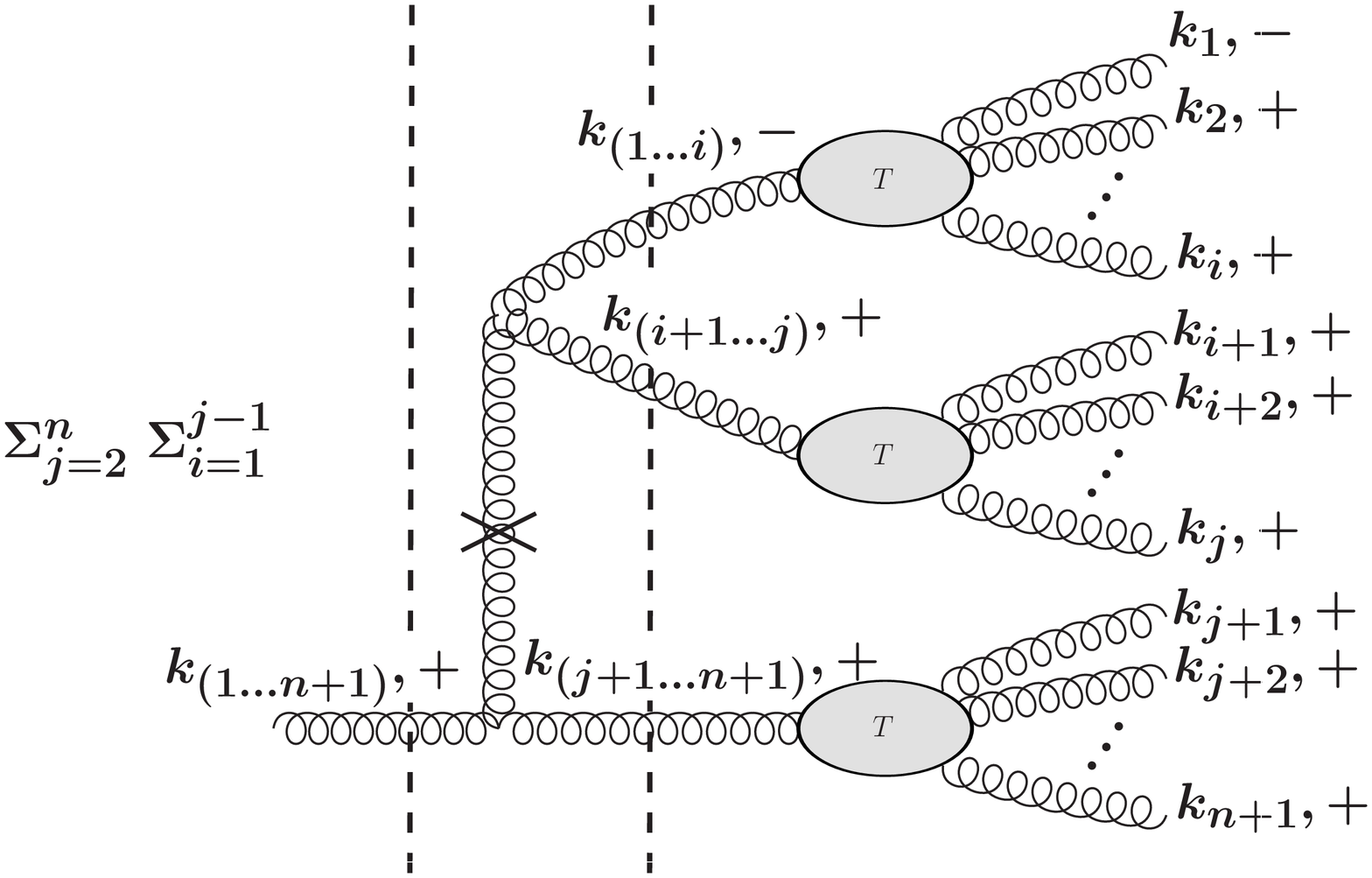}}}
\caption{Graphs involved in the fragmentation of a single off-shell gluon into $n+1$ on-shell gluons.  The initial and final helicities are specified in the figures.  We denote the 3-gluon vertex in Figs. \ref{fig:recurgrapha} and  \ref{fig:recurgraphb} as $V_+$ and $V_-$ respectively, the 4-gluon vertex in Fig.  \ref{fig:recurgraphc} as $V_4$, and the Coulomb term in Fig.  \ref{fig:recurgraphd} as $V_\text{Coul}$.
Vertical lines denote the energy denominators that need to be taken, they are implicit in all intermediate states denoted by blobs. There are no energy denominators in the final state.}
\label{fig:recurgraphs}
\end{figure}

 The above relation is illustrated in Fig.~\ref{fig:aandm} and was discussed at length in previous work \cite{CruzSantiago:2013wz} and used to compute the low order scattering amplitudes from the gluon wave functions.
 It turns out that in order to find ${\cal A}_{2\to n}$  or equivalently  $M_{1\to n+1}$ one actually needs to work with the different object, 
 $\overline{M}_{1\to n+1}$, which we define as the off-shell amplitude. This is obtained  by assuming the incoming gluons are off-shell and it is similar to the Berends-Giele recursion relations which involve the currents $J^{\mu}$ necessary for evaluation of the on-shell amplitudes. $\overline{M}_{1\to n+1}$ is more general than $M_{1\to n+1}$ in the sense that
\be
M_{1\to n+1} = \frac{1}{{\cal N}} \; \overline{M}_{1\to n+1}|_{D_{n+1}\to0} \; ,
\ee
where $D_{n+1}$ is the energy denominator for the first state. In the following, we will concentrate on  a particular configuration of helicities, $\overline{M}_{1\to n+1}(+ \,\rightarrow \, -+ \dots +)$, which corresponds to transition of $2\rightarrow n$ particles  for $(++\,\rightarrow \, +\dots +)$ when the two particles are incoming and the rest is outgoing or to scattering amplitude $(--+\dots+)$ when all the particles are outgoing. A key component in this derivation will be the fragmentation functions as defined in \cite{ms}, since it allows us to write $\overline{M}_{1\to n+1}$ as the sum of all the graphs in Fig. \ref{fig:recurgraphs}.  The convenience comes from the use of the factorization property of fragmentation functions.  
One can  write,
\begin{align}
\overline{M}_{1\to n+1}=&\, \sum _{j=2} ^n V_+
 \sqrt{\frac{z_1z_2\ldots z_{n+1}}{z_{1\ldots j}z_{j+1\ldots n+1}}}\; 
T_j[(1\ldots j)^+ \to 1^-,2^+,\ldots,j^+\,] \; T_{n+1-j}[(j+1\ldots n+1)^+ \to( j+1)^+,\ldots,(n+1)^+] \nonumber \\
&\,+\sum _{j=1} ^n V_-
 \sqrt{\frac{z_1z_2\ldots z_{n+1}}{z_{1\ldots j}z_{j+1\ldots n+1}}}\;  
T_j[(1\ldots j)^- \to 1^-,2^+,\ldots,j^+\,] \; T_{n+1-j}[(j+1\ldots n+1)^+ \to (j+1)^+,\ldots,(n+1)^+] \nonumber \\
&\,+\sum _{j=2} ^{n} \sum_{i=1}^{j-1} (V_4+V_\text{Coul})\;  
 \sqrt{\frac{z_1z_2\ldots z_{n+1}}{z_{1\ldots i}z_{i+1\ldots j}z_{j+1\ldots n+1}}}
T_i[(1\ldots i)^- \to 1^-,2^+,\ldots,i^+\,] \; \nonumber \\
&\quad \quad \quad \quad \quad \times \;  T_{j-i}[(i+1\ldots j)^+ \to (i+1)^+,\ldots,j^+] \; T_{n+1-j}[(j+1\ldots n+1)^+ \to (j+1)^+,\ldots,(n+1)^+]  \; .
\label{eq:recur}
\end{align}
The first, second and third line come from Fig. \ref{fig:recurgrapha}, Fig. \ref{fig:recurgraphb} and Figs. \ref{fig:recurgraphc} -  \ref{fig:recurgraphd} respectively.  The $V$'s are the vertex factors and these are given by
\be
V_+=2gz_{1\ldots n+1}v_{(j+1\ldots n+1)(1\ldots j)}^* \; , 
\ee
\be
V_-=2gz_{j+1\ldots n+1}v_{(1\ldots n+1)(1\ldots j)} \; ,
\ee
\be
V_4 =ig^2 \; ,
\ee
\be
V_{\text{Coul}} = ig^2 \frac{(z_{1\ldots n+1}+z_{j+1\ldots n+1})(z_{i+1\ldots j}-z_{1\ldots i})}{(z_{1\ldots n+1}-z_{j+1\ldots n+1})^2} \; .
\ee

Inspecting formula \eqref{eq:recur} we see that the  fragmentation functions involved in the process correspond to three different helicity configurations.  One of them $T_n[(12\ldots n)^+ \to 1^+,2^+,\ldots, n^+]$ was found in \cite{ms}  and its explicit expression was given in Eq.~\ref{eq:frag1}. The second one can be easily  derived using similar methods (see Appendix A) with the result

\be
T_n[(12\ldots n)^- \to 1^-,2^+,\ldots, n^+]=(-ig)^{n-1}\left(\frac{z_1}{z_{1\ldots n}}\right)^2\left(\frac{z_{1\ldots n }}{z_1\ldots z_n}\right)^{3/2} \frac{1}{v_{n\;n-1}v_{n-1\;n-2}\ldots v_{21}} \; .
\ee 
The third fragmentation function, $T_n[(12\ldots n)^+ \to 1^-,2^+,\ldots, n^+]$, however, remains unknown.  To find it we would, once again, need the graphs depicted in Fig. \ref{fig:recurgraphs}.  This implies a relationship between $\overline{M}_{1\to n}$ and $T_n[(12\ldots n)^+ \to 1^-,2^+,\ldots, n^+]$ which one can express as
\be
T_n[(12\ldots n)^+\to 1^-,2^+,\ldots, n^+] = \frac{1}{\sqrt{z_{1\ldots n} z_1 \ldots z_n}} \frac{i}{D_n} \overline{M}_{1\to n} \; .
\label{eq:Tsol}
\ee
Therefore this fragmentation function is directly proportional to $\overline{M}_{1\rightarrow n}$,
but it includes the denominator for the first (leftmost state) and different normalization of the external particles.
Thus, Eq.~\eqref{eq:recur} which is depicted in Fig.~\ref{fig:recurgraphs} turns out to be a recursion relation for $\overline{M}_{1\to n+1}$.  In the next section we will find a solution to this equation and prove it via the method of mathematical induction.

%%%%%%%%%%%%%%%%%%%%%%%%%%%%%%%%%%%%
%%%%%%%%%%%%%%%%%%%%%%%%%%%%%%%%%%%%%
\section{Explicit solution to the recursion formula}

In this section we will solve the recursion formula given in Eq.~\ref{eq:recur} supplemented by relation \eqref{eq:Tsol} and find the general expression for the off-shell amplitude $\overline{M}_{1\to n+1}$. We shall see that it can be expressed as a linear combination of the on-shell amplitudes with different number of external legs. By putting on-shell constraint on the external gluons, that is by putting $D_{n+1}=0$,  the on-shell MHV amplitude will be reproduced from the solution.

Let us start with the initial conditions for the recursion formula. The normalization is such that the initial fragmentation functions are set to 
 $T_1[1^+\to1^-]=0, \; T_1[1^+\to1^+]=T_1[1^-\to1^-]=1$.
Finding $\overline{M}_{1 \to 2}$ ($n=1$) is trivial,
\be
\overline{M}_{1\to 2}= 2gz_{2}v_{(1 2)1} \sqrt{\frac{z_1z_2}{z_{1}z_{2}}} 
= 2gz_{2}v_{(1 2)1} \frac{z_{12}^2v_{(1 2)1}v_{(1 2)1}}{z_{12}^2v_{(1 2)1}v_{(1 2)1}}
= 2g\frac{z_{1}z_{12}^2}{z_2} \frac{v_{(1 2)1}^3}{z_1v_{21} v_{21}}
= 2g\frac{z_{1}z_{12}}{z_2} \frac{v_{(1 2)1}^3}{v_{21} v_{2(12)}} 
=  2 i^2g^1 M_{1\to 2}\; ,
\ee
where we defined 

\be
M_{1\to 2}= \frac{z_{1}z_{12}}{z_2} \frac{v_{(1 2)1}^3}{v_{21} v_{2(12)}}  \; .
\ee

\begin{figure}[h]
\centering
\subfloat[]{\label{fig:recurgraphe}\includegraphics[width=.488\textwidth]{{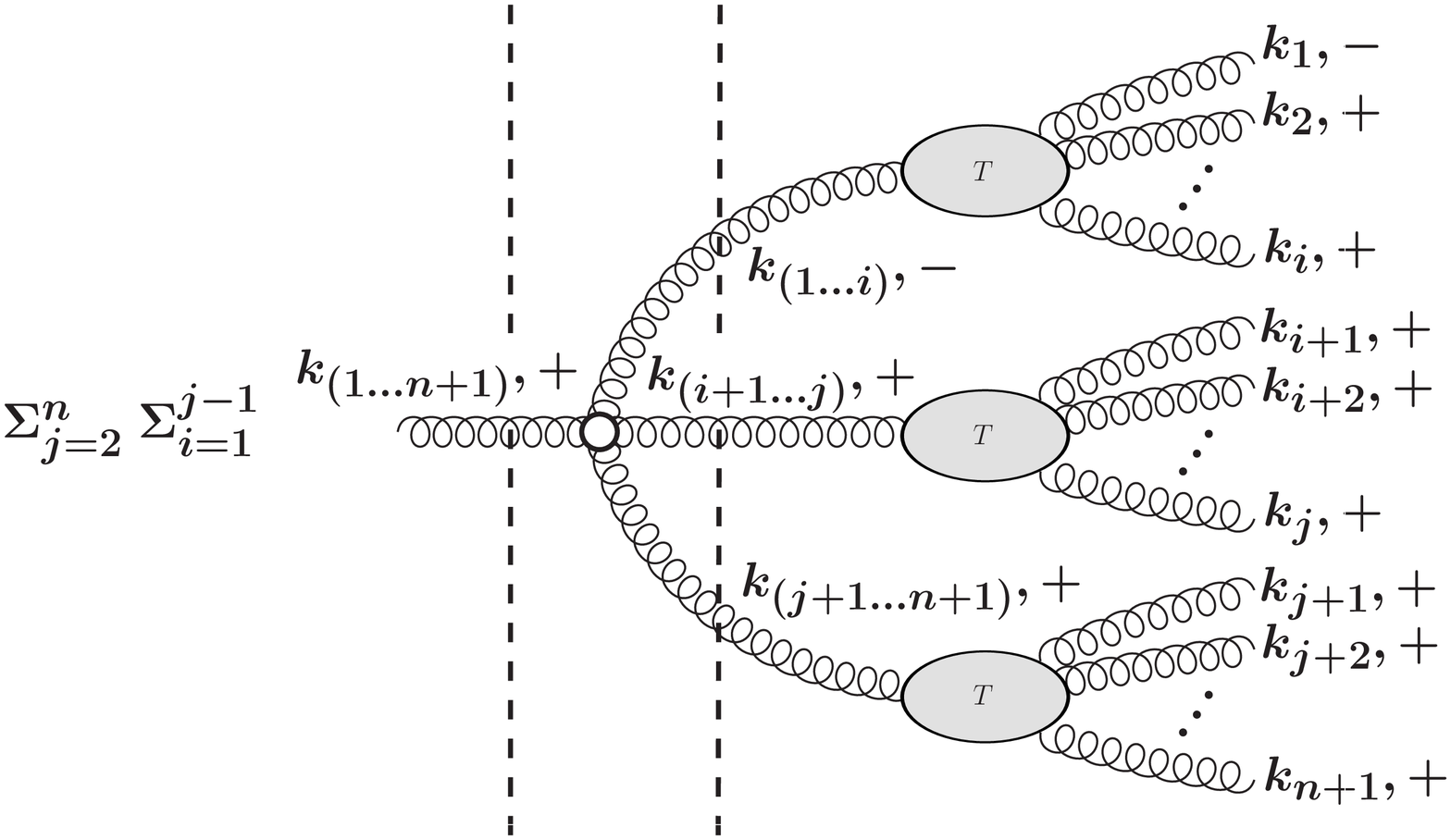}}} \;\;\;
\subfloat[]{\label{fig:recurgraphf}\includegraphics[width=.488\textwidth]{{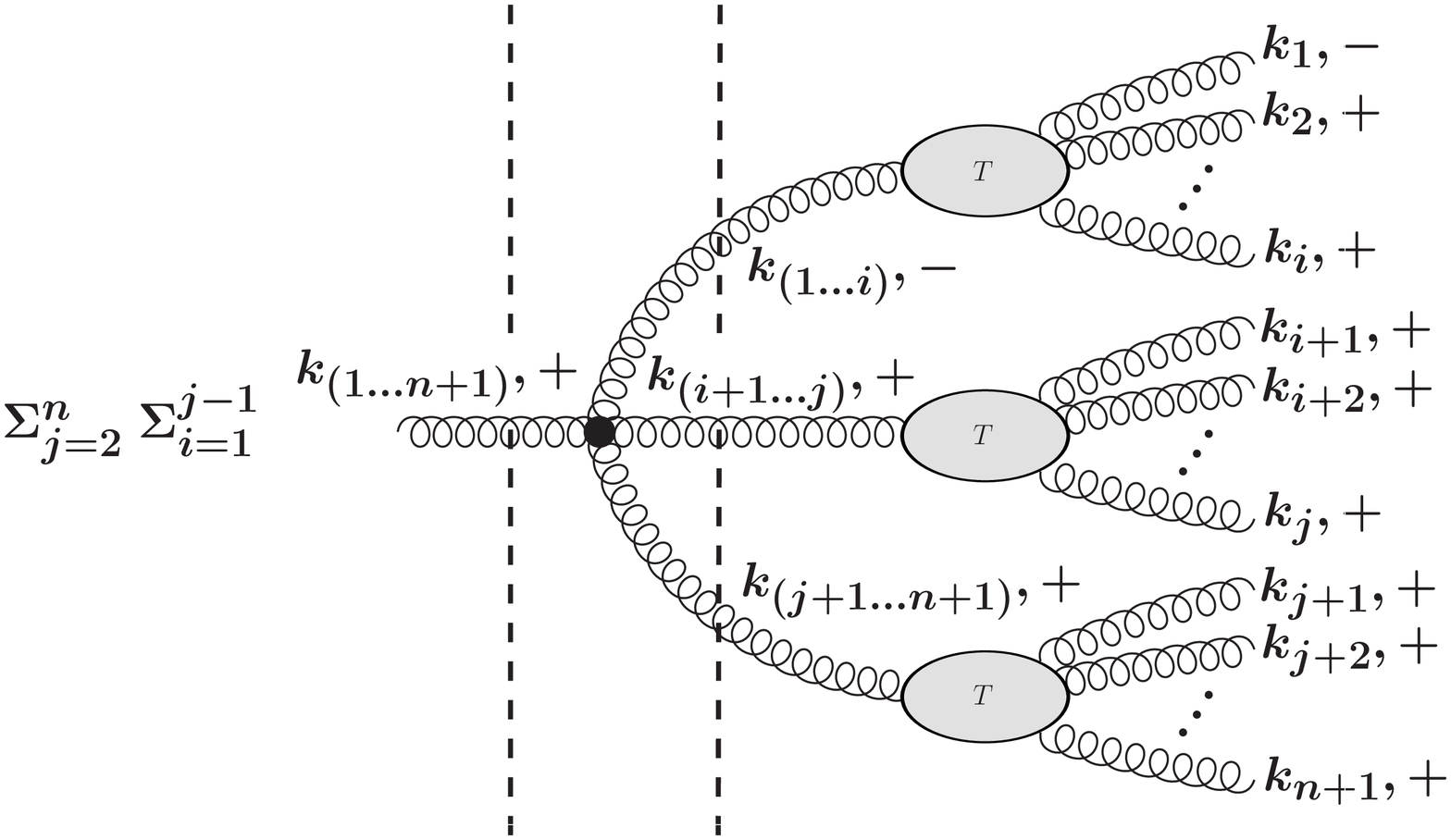}}} 
\caption{Graphs representing the contribution from the equivalent 4-gluon vertices $ V_\text{comb, a}$ and $ V_\text{comb, b}$ to the fragmentation function.}
\label{fig:new4vertexgraphs}
\end{figure}

 Finding $\overline{M}_{1 \to 3}$ ($n=2$) is much more complicated and we should remark that the order in which terms will be added is the same as when we perform the proof via induction.  The order is important because it makes the structure of the solution easier to see.
To begin, it will be convenient for our calculations to add $V_4$ and $V_\text{Coul}$ to get
\be
V_4+V_\text{Coul} = V_\text{comb, a}+V_\text{comb, b} \; ,
\ee
where
\be
 V_\text{comb, a}=2ig^2\frac{z_{1\ldots n+1} \;z_{i+1\ldots j}}{z_{1\ldots j}^2} \; ,
\ee
\be
 V_\text{comb, b}=-2ig^2\frac{z_{1\ldots i}\; z_{j+1\ldots n+1}}{z_{1\ldots j}^2} \; .
\ee
Thus, we can replace Figs. \ref{fig:recurgraphc} and \ref{fig:recurgraphd}  with Figs.  \ref{fig:recurgraphe} and \ref{fig:recurgraphf}.  The white and black blobs represent the contributions from the vertices $V_\text{comb, a}$ and $V_\text{comb, b}$ respectively.
From recursion  \eqref{eq:recur} we see that there are  five different terms which contribute to $\overline{M}_{1\to 3}$:
\be
\text{I} = 2gz_{123}v_{(3)(12)}^* \sqrt{\frac{z_1z_2 z_3}{z_{12}z_{3}}} \left[\frac{1}{\sqrt{z_{12} z_1 z_2}} \frac{i}{D_2} \overline{M}_{1\to 2} \right] \; ,
\ee
\be
\text{II} = 2gz_{23}v_{(123)1} \sqrt{\frac{z_1z_2 z_{3}}{z_{1}z_{23}}} \left[(-ig)^{1}\left(\frac{z_{23 }}{z_2 z_3}\right)^{3/2} \frac{1}{v_{32}}\right] = -2ig^2\frac{z_{23}^2}{z_2z_3}\frac{v_{(321)1}}{v_{32}} \; ,
\ee
\be
\text{III} = 2gz_{3}v_{(123)(12)} \sqrt{\frac{z_1z_2 z_{3}}{z_{12}z_{3}}} \left[(-ig)^{1}\left(\frac{z_1}{z_{12}}\right)^2\left(\frac{z_{12 }}{z_1 z_2}\right)^{3/2} \frac{1}{v_{21}}\right] 
= -2ig^2\frac{z_1z_3}{z_{12}z_2}\frac{v_{(321)(21)}}{v_{21}} \; ,
\ee
\be
\text{IV} = 2ig^2\frac{z_{123} \;z_{2}}{z_{12}^2}  \sqrt{\frac{z_1z_2 z_3}{z_{1}z_{2}z_{3}}} 
= -2ig^2 \frac{z_{123}z_3}{z_{12}^2 z_3} \frac{M_{1\to 2}}{v_{(12)1}} \; ,
\ee
\be
 \text{V} = -2ig^2\frac{z_{1}\; z_{3}}{z_{12}^2}\sqrt{\frac{z_1z_2 z_3}{z_{1}z_{2}z_{3}}} \; .
\ee
Here we have already written IV in terms of ${M}_{1\to 2}$ as a simple example of how the term coming from Fig. \ref{fig:recurgraphe}  will be simplified later on. Next, III is added to V to get
\begin{align}
\text{VI} &= \text{III} + \text{V}=-2ig^2\frac{z_1z_3}{z_{z12}}\left[\frac{1}{z_{12}}+\frac{1}{z_2}\frac{v_{(321)(21)}}{v_{21}}\right]
=-2ig^2\frac{z_1z_3}{z_2z_{12}}\frac{v_{(123)1}}{v_{21}} \; ,
\end{align}
which added to II gives
\be
\text{VI}+\text{II}=\text{VII} =  -2ig^2\frac{v_{(123)1}}{z_2}\left[\frac{z_{23}^2}{z_3}\frac{1}{v_{32}} +\frac{z_1z_3}{z_{12}}\frac{1}{v_{21}}\right] \; .
\ee
However, using the following relation
\begin{align}
\frac{z_{123}z_2}{z_{12}}+\frac{z_1z_3^2}{z_{12}z_2}\frac{v_{3(123)}}{v_{12}}+\frac{z_{23}^2}{z_2}\frac{v_{3{(123)}}}{v_{23}}&=\frac{1}{z_{12}z_2}\frac{1}{v_{12}v_{23}}\left[z_{123}z_2^2v_{12}v_{23}+z_1z_3^2v_{3{(123)}}v_{23}+z_{12}z_{23}^2v_{3({123})}v_{12}\right]\\
&=\frac{1}{z_{12}z_2}\frac{1}{v_{12}v_{23}}\left[z_{123}z_2v_{12}\left(z_2v_{23}+z_{23}v_{3({123})}\right)+z_1z_3v_{3({123})}\left(z_3v_{23}+z_{23}v_{12}\right)\right]\\\\
&=\frac{z_{123}z_1}{z_{12}z_2}\frac{v_{({123})1}}{v_{12}v_{23}}\left[-z_{12}v_{({123})1}\right]  \; ,
\label{prop2}
\end{align}
we see that  term VII can be written as
\begin{align}
\text{VII}&= -2ig^2 v_{(123)1}
 \left\{\frac{M_{1\to 3}}{v_{(123)1}} -  \frac{1}{z_3} \frac{z_{123}^2}{z_{12}z_{123}} \frac{1}{v_{3(123)}}\frac{M_{1\to 2}}{v_{(12) 1}} \right\}  \; .
\end{align}
Now that $\overline{M}_{1\to 3}$ = I + IV + VII is written completely in terms of on-shell amplitudes we can collect terms proportional to $M_{1\to j}$ to get, 
\begin{align}
\overline{M}_{1\to 3} &= -2ig^2 M_{1\to 3} +2ig^2 \frac{z_{123}}{ z_{12}} M_{1\to 2}\left(\frac{1}{z_3}\frac{v_{(123)1}}{v_{3(123)}v_{(12)1}} -\frac{1}{z_{12}v_{(12)1}}-2\frac{v_{3(12)}^*}{D_2}\right) \\ \nonumber\\
&=-2ig^2 M_{1\to 3} -2ig^2 \frac{1}{z_3}\frac{z_{123}}{z_{12}^2 z_{123}}\frac{1}{v_{3(123)}}\frac{M_{1\to 2}}{v_{(12)1}D_2} \\
& \times \, \left\{D_2\left(-z_{123}z_{12}v_{(123)1}+z_3 z_{123} v_{3(123)}\right)  +z_{12}v_{(12)1} \left(2z_3 z_{123} v_{3(12)}^* v_{3(123)}\right)\right\} \; .
\end{align}
Using
\begin{align}
2z_3 z_{123} v_{3(12)}^* v_{3(123)} = -z_3z_{12}\left(\frac{\uvec{k}_3}{z_3}-\frac{\uvec{k}_{12}}{z_{12}}\right)^2
= z_{123}(D_2 - D_3) \; ,
\end{align}
and
\be
-z_{123}z_{12}v_{(123)1}+z_3 z_{123} v_{3(123)}=-z_{123}z_{12}v_{(123)1}-z_{12} z_{123} v_{(12)(123)}=-z_{123}z_{12}v_{(12)1} \; ,
\ee
our final result for $\overline{M}_{1\to 3}$ is then
\be
\overline{M}_{1\to 3}=2i^3g^2 M_{1\to 3} -2i^3g^2 \frac{1}{z_3}\frac{z_{123}}{z_{12}}\frac{1}{v_{3(123)}}\frac{D_3}{D_2}M_{1\to 2} \;  ,
\ee
where $M_{1\to 3}$ is a special case  of the general definition for arbitrary number $n$ of particles
\be
M_{1\to n}\equiv \frac{z_{1\ldots n}\;z_1}{z_2 z_3\ldots z_n} \frac{v_{(1\ldots n)1}^3}{v_{12}v_{23}\ldots v_{n-1\;n}v_{n(1\ldots n)}}\;,
\label{eq:Mdef}
\ee
which (up to some factors) is the on-shell scattering amplitude for $2\to n-1$ transition,
 ${\cal A}_{2\to n-1}$, see Eq.~\ref{eq:aandm}.

The next step in the iteration, $\overline{M}_{1\to 4}$, can be found following the same procedure, yet it is a much more tedious process.  The result ends up being
\begin{align}
\overline{M}_{1\to4} = 2 i^4  g^3 \left\{M_{1 \to 4} - \frac{D_4}{D_3} \frac{z_{1234}^2}{z_{123}z_{1234}}\frac{1}{z_4} \frac{1}{v_{4(1234)}} M_{1 \to 3} - \frac{D_4}{D_2}\frac{z_{1234}^2}{z_{12}z_{123}}\frac{1}{z_3 z_4} \frac{1}{v_{34}}\frac{M_{1\to 2}}{v_{3(123)}} \right\} \; .
\end {align}
Interestingly, the off-shell amplitude is expressed as a linear combination of the on-shell objects with the pre factors which are proportional to the energy denominators.
In particular we see that by putting the on-shell constraint $D_4=0$ we recover the on-shell amplitude.

Following the pattern found, one would then expect, for a general integer $n\ge2$, 
\be
\overline{M}_{1\to n} = 2 i^n g^{n-1} \left\{M_{1\to n} - z_{1\ldots n}^2 D_n \sum_{i=2}^{n-1} \frac{1}{z_{1\ldots i}z_{1\ldots i+1}} \frac{1}{z_{i+1} \ldots z_n} \frac{1}{v_{i+1\;i+2} \ldots v_{n-1\;n}}\frac{M_{1\to i}}{v_{i+1(1\ldots i+1)} D_i} \right\} \; .
\label{eq:Msol}
\ee
We shall now present the  proof of this result using the method of mathematical induction.  Before we begin, the following are two relationships which we will use many times in the rest of this paper,
\be
 z_{1\ldots j+1}v_{(1\ldots j+1)1}=z_{2\ldots j+1}v_{(2\ldots j+1)1} = \sum_{i=1}^{j}z_{i+1\ldots j+1} v_{i+1\;i} \; ,
 \label{eq:rel1} 
\ee
and
\be
 z_{i+1\ldots j+1}v_{j+1(i+1\ldots j+1)}=z_{i+1\ldots j}v_{j+1(i+1\ldots j)} = \sum_{l=i+1}^{j} z_{i+1\ldots l} v_{l+1\;l} \; .
\label{eq:rel2}
 \ee
%%%%%%%%%%%%%%%%%%%%%%%%%%%%%%%%%%%%
%%%%%%%%%%%%%%%%%%%%%%%%%%%%%%%%%%%%
To perform the proof we assume \eqref{eq:Msol} is true and then use it in \eqref{eq:recur} to find $\overline{M}_{n+1}$.  At the end, $\overline{M}_{n+1}$ should be of the form given by expression \eqref{eq:Msol} for $n\to n+1$.   Let us remind that  for the result one needs to add all the contributions from Figs. \ref{fig:recurgrapha}, \ref{fig:recurgraphb}, \ref{fig:recurgraphe} and \ref{fig:recurgraphf}.   We begin with Fig. \ref{fig:recurgraphe}. For fixed $j$, the expression for this graph reads
%%%%%%%%%%%%
\begin{align}
{\cal E}_1=-\sum_{i=1}^{j-1} 2 (-i)^{n-1}g^n \frac{z_{1\ldots n+1}z_1 z_{j+1\ldots n+1}z_{i+1 \ldots j}^2}{z_{1\ldots i} z_{1\ldots j}^2\; z_2\ldots z_{n+1}}\frac{1}{v_{n+1\;n}\ldots v_{j+2\;j+1}v_{j\;j-1}\ldots v_{i+2\;i+1} v_{i\;i-1}\ldots v_{21}}\nonumber \\ 
=-2 (-i)^{n-1}g^n \frac{z_{1\ldots n+1} z_{j+1\ldots n+1}}{z_{1\ldots j}^2\; z_2\ldots z_{n+1}}\frac{1}{v_{n+1\;n}\ldots v_{j+2\;j+1}} 
\sum_{i=1}^{j-1} \frac{z_1z_{i+1\ldots j}^2}{z_{1\ldots i}}\frac{1}{v_{j\;j-1}\ldots v_{i+2\;i+1}v_{i\;i-1}\ldots v_{21}} \; .
\label{eq:4v1}
\end{align}
%%%%%%%%%%%%
Next, we add the expressions for the  graphs presented  in Figs. \ref{fig:recurgraphb} and \ref{fig:recurgraphf} for fixed $j$ ($j\ne1$),
%%%%%%%%%%%%
\begin{align}
A_j&=\;2(-i)^{n-1}g^n \frac{z_1z_{j+1\ldots n+1}^2}{z_{1\ldots j}\;z_2\ldots z_{n+1}}\frac{v_{(1\ldots n+1)1}}{v_{n+1\;n}\ldots v_{j+2\;j+1}v_{j\;j-1}\ldots v_{21}} \; ,
\end{align}
where we have used Eq.~\eqref{eq:rel1} .
%%%%%%%%%%%%
For $j=1$ in Fig. \ref{fig:recurgraphb},
%%%%%%%%%%%%
\begin{align}
B&=2(-i)^{n-1}g^n\frac{z_{2\ldots n+1}^2}{z_2\ldots z_{n+1}}\frac{v_{(1\ldots n+1)1}}{v_{n+1\;n}\ldots v_{32}} \; .
\end{align}
%%%%%%%%%%%%
The overall contribution from Figs. \ref{fig:recurgraphb} and \ref{fig:recurgraphf} would then be given by
%%%%%%%%%%%%
\begin{align}
{\cal E}_2=\sum_{j=2}^n A_j+B = 2(-i)^{n-1}g^n\frac{v_{(1\ldots n+1)1}}{z_2\ldots z_{n+1}}
\sum_{j=1}^n  \frac{z_1z_{j+1\ldots n+1}^2}{z_{1\ldots j}}\frac{1}{v_{n+1\;n}\ldots v_{j+2\;j+1}v_{j\;j-1}\ldots v_{21}}\; .
\label{eq:4v2}
\end{align}
%%%%%%%%%%%%
For Fig. \ref{fig:recurgrapha}, for fixed $j$, we get
%%%%%%%%%%%%
\begin{align}
{\cal E}_3=2 (-1)^{n-j}(ig)^{n-j+1} \frac{  z_{1\ldots n+1} z_{j+1\ldots n+1}}{z_{1\ldots j}z_{j+1}\ldots z_{n+1}}   \frac{  v_{(n+1\ldots j+1)(j\ldots 1)}^*}{v_{n+1\;n}\ldots v_{j+2\;j+1}}
 \frac{1}{D_j} \overline{M}_{1\to j} \;  . 
\label{eq:finalexp1}
\end{align}
%%%%%%%%%%%%
Now, we need to add the contributions from \eqref{eq:4v1}, \eqref{eq:4v2} and \eqref{eq:finalexp1} to get the final expression for $\overline{M}_{n+1}$.  To simplify the calculations it is useful to rewrite the expressions  entirely in terms $M_{1\to j}$.  We can use the following identity (which is proven in Appendix B)
\begin{multline}
\frac{v_{k+1\;(1\ldots k+1)}}{z_2 z_3 \ldots z_k} \sum_{j=1}^k  \frac{z_1z_{j+1\ldots k+1}^2}{z_{1\ldots j}}\frac{1}{v_{k+1\;k}\ldots v_{j+2\;j+1}v_{j\;j-1}\ldots v_{21}}\\
=(-1)^k z_{k+1} v_{k+1(1\ldots k+1)}\left\{\frac{M_{1\to k+1}}{v_{(1\ldots k+1)1}} - \sum_{j=2}^{k} \frac{1}{z_{j+1}\ldots z_{k+1}} \frac{z_{1\ldots k+1}^2}{z_{1\ldots j}z_{1\ldots j+1}} \frac{1}{v_{j+1\; j+2} \ldots v_{k\;k+1}} \frac{1}{v_{j+1(1\ldots j+1)}}\frac{M_{1\to j}}{v_{(1\ldots j) 1}} \right\} \; ,
\label{eq:relation}
\end{multline}
to rewrite ${\cal E}_1$ in \eqref{eq:4v1} and ${\cal E}_2$ in \eqref{eq:4v2},

\begin{align}
{\cal E}_1  & =2 (-i)^{n-1}g^n \frac{z_{1\ldots n+1}
 z_{j+1\ldots n+1}}{z_{1\ldots j}^2\; z_{j+1}\ldots z_{n+1}}\frac{1}{v_{n+1\;n}\ldots v_{j+2\;j+1}}  \\
& \times \, (-1)^{j} \left\{\frac{M_{1\to j}}{v_{(1\ldots j)1}} - \sum_{i=2}^{j-1} \frac{1}{z_{i+1}\ldots z_{j}} \frac{z_{1\ldots j}^2}{z_{1\ldots i}z_{1\ldots i+1}} \frac{1}{v_{i+1\; i+2} \ldots v_{j-1\;j}} \frac{1}{v_{i+1(1\ldots i+1)}}\frac{M_{1\to i}}{v_{(1\ldots i) 1}} \right\} \; ,
\label{eq:finalexp2}
\end{align}

\begin{align}
{\cal E}_2&= -2(i)^{n-1}g^n v_{(1\ldots n+1)1}
 \left\{\frac{M_{1\to n+1}}{v_{(1\ldots n+1)1}} - \sum_{j=2}^{n} \frac{1}{z_{j+1}\ldots z_{n+1}} \frac{z_{1\ldots n+1}^2}{z_{1\ldots j}z_{1\ldots j+1}} \frac{1}{v_{j+1\; j+2} \ldots v_{n\;n+1}} \frac{1}{v_{j+1(1\ldots j+1)}}\frac{M_{1\to j}}{v_{(1\ldots j) 1}} \right\} \; .
 \label{eq:finalexp3}
\end{align}

We can now find $\overline{M}_{1\to n+1}$ from the contributions of \eqref{eq:finalexp1}, \eqref{eq:finalexp2} and \eqref{eq:finalexp3}, where we must remember to sum over $j$ in \eqref{eq:finalexp1} and \eqref{eq:finalexp2}  from $j=2$ to $j=n$ and collect terms proportional to $M_{1\to l}$, where $2\le l\le n+1$.  For $l=n+1$ we get only one term, which comes from the first term in \eqref{eq:finalexp3},
\be
2i^{n+1}g^n M_{1\to n+1} \; .
\label{eq:firstterm}
\ee
We should note that this term is, in fact, the MHV amplitude, ${\cal A}_{2\to n}$, we wished to obtain.
For any other $l$, after simplifying and remembering to use \eqref{eq:Msol} we get 
\begin{align}
&2i^{n+1} g^n \frac{1}{z_{l+1}\ldots z_{n+1}}\frac{z_{1\ldots n+1}}{z_{1\ldots l}^2 z_{1\ldots l+1}} \frac{1}{v_{l+1\; l+2} \ldots v_{n\;n+1}} \frac{1}{v_{l+1(1\ldots l+1)}}\frac{M_{1\to l}}{v_{(1\ldots l) 1} D_l}\left\{D_l C+  z_{1\ldots l} v_{(1\ldots l)1}F \right\} \; ,
\label{eq:D_simp}
\end{align}
where
\be
C=- z_{1\ldots n+1} z_{1\ldots l} v_{(1\ldots n+1)1}   +z_{l+1\ldots n+1} z_{1\ldots l+1} v_{l+1(1\ldots l+1)} - \sum_{j=l+1}^n z_{j+1\ldots n+1} z_{1\ldots l} v_{j\;j+1}  \; ,
\label{eq:Mterm1}
\ee
\be
F=2 z_{l+1\ldots n+1} z_{1\ldots l+1} v^*_{(n+1\ldots l+1)(l\ldots 1)} v_{l+1(1\ldots l+1)}  -2\sum_{j=l+1}^n z_{j+1\ldots n+1} z_{1\ldots j} v^*_{(n+1\ldots j+1)(j\ldots 1)} v_{j\;j+1} \; .
\ee
However, we can write $C$ as
\begin{align}
C&=-z_{1\ldots l}^2 v_{(1\ldots l)1} -\left(z_{1\ldots l} \sum_{j=l+1}^{n+1} z_j v_{j\;1} \right) +z_{1\ldots l}  \left(\sum_{j=l+1}^{n+1} z_j \right) v_{l+1(1\ldots l)}  -z_{1\ldots l} \sum_{j=l+1}^n \sum_{m=j+1}^{n+1} z_m v_{j\;j+1} \; .
\end{align}
After some little algebra and changing the order of summation in the last term, i.e. replacing $\sum_{j=l+1}^n \sum_{m=j+1}^{n+1}$ with $\sum_{m=l+2}^{n+1} \sum_{j=l+1}^{m-1}$,  we arrive at
\begin{align}
C&= - z_{1\ldots l} z_{1\ldots n+1} v_{(1\ldots l) 1} \; .
\end{align}
Furthermore, we can write $F$ as
\begin{align}
F&=2 \sum_{j=l+1}^{n+1}z_{j}  z_{1\ldots l} v^*_{j(l\ldots 1)} v_{l+1(1\ldots l)}
+2\sum_{j=l+1}^n \sum_{m=j+1}^{n+1}z_{m} z_{1\ldots j} v^*_{m(j\ldots 1)} v_{j+1\;j} \; .  
\end{align}
Since $j$ will always be greater than $l$ we can write $z_{1\ldots j} v^*_{m(j\ldots 1)} =  z_{1\ldots l} v^*_{m(l\ldots 1)} +\sum_{j=l+1}^j z_{i} v^*_{m\;i} $.  We use this to expand the second term in $F$.  Changing the order of summation in these two new terms we get
\begin{align}
F&=2 \sum_{j=l+1}^{n+1}z_{j}  z_{1\ldots l} v^*_{j(l\ldots 1)} v_{l+1(1\ldots l)}
+2z_{1\ldots l}\sum_{m=l+2}^{n+1} z_{m}  v^*_{m(l\ldots 1)} v_{m\;l+1} 
+2\sum_{m=l+2}^{n+1}\sum_{i=l+1}^{m-1}z_{m} z_{i} v^*_{m\;i} v_{m\;i} \\
&= 2 \sum_{j=l+1}^{n+1}z_{j}  z_{1\ldots l} v^*_{j(l\ldots 1)} v_{j(1\ldots l)}
+2\sum_{j=l+2}^{n+1}\sum_{i=l+1}^{j-1}z_{j} z_{i} v^*_{j\;i} v_{j\;i}  \; .
\end{align}
We can rewrite this in terms of $k$'s by using
\be
v^*_{ab}v_{ab} \;  = \;  -\frac{1}{2} \left(\frac{\uvec{k}_a}{z_a}-\frac{\uvec{k}_b}{z_b}\right)^2 \, .
\ee
Manipulating the sums and simplifying the expression we end up with
\begin{align}
F&=- z_{1\ldots n+1}\left(\frac{\uvec{k}_{1\ldots l}^2}{z_{1\ldots l}} +\sum_{j=l+1}^{n+1}\frac{\uvec{k}_j^2}{z_j}\right)+\left(\uvec{k}_{1\ldots l}^2 +  \sum_{j=l+1}^{n+1} \uvec{k}^2_{j}\right)+ 2\uvec{k}_{l+1\ldots n+1}\cdot \uvec{k}_{1\ldots l}+\sum_{j=l+2}^{n+1}\sum_{i=l+1}^{j-1} 2\uvec{k}_j \cdot \uvec{k}_i \\
&=  z_{1\ldots n+1}\left(\frac{\uvec{k}_{1\ldots n+1}^2}{z_{1\ldots n+1}} - \frac{\uvec{k}_{1\ldots l}^2}{z_{1\ldots l}} -\sum_{j=l+1}^{n+1}\frac{\uvec{k}_j^2}{z_j}\right)  \\
&=  z_{1\ldots n+1}(D_l - D_{n+1}) \; .
\end{align}
Thus, \eqref{eq:D_simp} reduces to
\be
-2i^{n+1} g^n \frac{1}{z_{l+1}\ldots z_{n+1}}\frac{z_{1\ldots n+1}^2}{z_{1\ldots l}z_{1\ldots l+1}} \frac{1}{v_{l+1\; l+2} \ldots v_{n\;n+1}} \frac{1}{v_{l+1(1\ldots l+1)}}\frac{D_{n+1}}{ D_l}M_{1\to l} \; .
\label{eq:otherterms}
\ee
The final result is the sum of \eqref{eq:firstterm} and \eqref{eq:otherterms} 
\be
2i^{n+1}g^n M_{1\to n+1}-2i^{n+1} g^n \sum_{l=2}^n \frac{1}{z_{l+1}\ldots z_{n+1}}\frac{z_{1\ldots n+1}^2}{z_{1\ldots l}z_{1\ldots l+1}} \frac{1}{v_{l+1\; l+2} \ldots v_{n\;n+1}} \frac{1}{v_{l+1(1\ldots l+1)}}\frac{D_{n+1}}{ D_l}M_{1\to l} \; ,
\label{eq:finalresult}
\ee
which 
gives us $\overline{M}_{1\to n+1}$ written as a linear combination of on-shell amplitudes.  Comparing these terms to \eqref{eq:Msol} we see that, indeed, $\overline{M}_{1\to n+1}$ is of the   same form, which completes our proof.  It is important to note that if we now apply the condition that the initial state be on-shell, i.e. $D_{n+1}\to0$,  $\overline{M}_{1\to n+1}$ does reduce to the known  expression for the MHV amplitudes.

%%%%%%%%%%%%%%%%
\section*{Conclusions}

In this paper we have investigated the  fragmentation functions and scattering amplitudes within the framework of the light front perturbation theory. 
We have shown that the recursion relations for the fragmentation functions actually imply the recursion relations for the off-shell light-front scattering amplitudes.
These recursion relations are light-front analogs of the previously derived Berends-Giele recursion relations. 
Using these methods we have been able to reproduce the lowest order scattering amplitudes. Finally, it was shown that the recursion relations can be solved exactly to all orders in the number of external legs and thus compact expressions for the off-shell amplitudes have been derived. Interestingly, the expression for the off-shell amplitude can be expressed as a sum of terms proportional to the on-shell amplitudes multiplied by the appropriate light-front energy denominators. When the external gluons are put on-shell, the energy denominators vanish, and the first term in the sum reproduces the previously known results for the MHV amplitudes.
The light-front methods presented in this paper can be readily generalized  to compute the scattering amplitudes  for different helicity configurations.

%%%%%%%%%%%%%%%%

\section*{Acknowledgments}

 This work was supported  in part  by the DOE OJI grant No. DE - SC0002145   and by  the   Polish NCN 
grant DEC-2011/01/B/ST2/03915.  A.M.S. is supported by the Sloan Foundation.

%%%%%%%%%%%%%%%%
\section*{Appendix A}
In this appendix we will show how the fragmentation function $T_n[(12\ldots n)^- \to 1^-,2^+,\ldots, n^+]$ is derived.  We will follow the procedure used  in \cite{ms} to calculate $T_n[(12\ldots n)^+ \to 1^+,2^+,\ldots, n^+]$.

We begin with $n=2$. It can easily be seen that the fragmentation function for this is
\be
T_2[(12)^- \to 1^-,2^+] = 2ig \frac{z_1}{\sqrt{z_{12}z_1 z_2}} \frac{v^*_{2(21)}}{D_2}
= 2ig \frac{z_1}{\sqrt{z_{12}z_1 z_2}} \frac{v^*_{2(21)}}{-2 \xi_{12}v_{21}^*v_{21}}
= -ig \left(\frac{z_1}{z_{12}}\right)^2 \left(\frac{z_{12}}{z_1z_2}\right)^{3/2} \frac{1}{v_{21}}.
\ee
One should note that this is the same as that for $T_2[(12)^+ \to 1^+,2^+]$ multiplied by an extra factor of $\left({z_1}/{z_{12}}\right)^2$. For n=3 we then get
\begin{align}
T_3[(123)^- \to 1^-,2^+,3^+] &= \frac{-2(ig)^2}{D_3}\left[\frac{z_{ 1} }{\sqrt{z_{123} z_{1} z_{23}}}\left(\frac{z_{23}}{z_2 z_3}\right)^{3/2} \frac{v_{(32)(321)}^*}{v_{32}} + \frac{z_{12} }{\sqrt{z_{123} z_{12} z_{3}}}\left(\frac{z_1}{z_{12}}\right)^2\left(\frac{z_{12}}{z_1 z_2}\right)^{3/2} \frac{v_{3(321)}^*}{v_{21}}\right]  \\
&=  \frac{-2(ig)^2}{D_3}\left[\left(\frac{z_{ 1} }{z_{123}}\right)^2\left(\frac{z_{23}}{z_2 z_3}\right)^{3/2} \frac{v_{(32)1}^*}{\sqrt{\xi_{1(23)}}v_{32}} + \left(\frac{z_{12} }{z_{123}}\right)^2\left(\frac{z_1}{z_{12}}\right)^2\left(\frac{z_{12}}{z_1 z_2}\right)^{3/2} \frac{v_{3(21)}^*}{\sqrt{\xi_{(12)3}}v_{21}}\right] \\
&=  \frac{-2(ig)^2}{D_3}\left(\frac{z_{ 1} }{z_{123}}\right)^2\left[ \frac{v_{(32)1}^*}{\xi_{23}^{3/2}\xi_{1(23)}^{1/2}v_{32}} + \frac{v_{3(21)}^*}{\xi_{12}^{3/2}\xi_{(12)3}^{1/2}v_{21}}\right].
\end{align} 
We recognize that this is the expression given in \cite{ms} for $T_3[(13)^+ \to 1^+,2^+,3^+]$ multiplied by $\left(z_1/z_{123}\right)^2$.  Thus, the final result for $n=3$ is 
\be
T_3[(123)^- \to 1^-,2^+,3^+] = (ig)^2 \left(\frac{z_{ 1} }{z_{123}}\right)^2 \left(\frac{z_{123}}{z_1 z_2 z_3}\right)^{3/2} \frac{1}{v_{32}v_{21}}.
\ee
This pattern implies that for general $n$ the fragmentation function should be of the form
\be
T_n[(12\ldots n)^- \to 1^-,2^+,\ldots, n^+] = \left(\frac{z_{1} }{z_{1\ldots n}}\right)^2 T_n[(12\ldots n)^+ \to 1^+,2^+,\ldots, n^+].
\label{eq:newfrag}
\ee
We show that this expression is indeed correct by substituting \eqref{eq:newfrag} into the recursion relation for $T_{n+1}[(12\ldots n+1)^- \to 1^-,2^+,\ldots, n+1^+]$ given below. 
\begin{multline}
T_{n+1}[(12 \ldots n+1)^- \to 1^-,2^+,\ldots,n+1^+] \; = \; -{2ig\over D_{n+1}} \; 
\sum _{i=1} ^n  \, \left\{\, \left(\frac{z_{1\ldots i}}{z_{1\ldots n+1}}\right)^2
{ v^*_{(1\ldots i)(i+1\ldots n+1)} 
\over \sqrt{\xi_{(1\ldots i)(i+1\ldots n+1)}}} \; \right. \\
\left. \times \;\rule{0em}{1.8em}
T_i[(1\ldots i)^- \to 1^-,2^+,\ldots,i^+\,] \; T_{n+1-i}[(i+1\ldots n+1)^+ \to i+1^+,\ldots,n+1^+]\,
\right\}\, . 
\label{eq:newfragrecur}
\end{multline}
We see that $T_i[(1\ldots i)^- \to 1^-,2^+,\ldots,i^+\,]$ includes a factor of $\left({z_{1} }/{z_{1\ldots i}}\right)^2$ which can be combined with the $\left({z_{1\ldots i}}/{z_{1\ldots n+1}}\right)^2$ in \eqref{eq:newfragrecur} to give 
\begin{multline}
T_{n+1}[(12 \ldots n+1)^- \to 1^-,2^+,\ldots,n+1^+] \; = \; -{2ig\over D_{n+1}}\left(\frac{z_{1}}{z_{1\ldots n+1}}\right)^2 \; 
\sum _{i=1} ^n  \, \left\{\, 
{ v^*_{(1\ldots i)(i+1\ldots n+1)} 
\over \sqrt{\xi_{(1\ldots i)(i+1\ldots n+1)}}} \; \right. \\
\left. \times \;\rule{0em}{1.8em}
T_i[(1\ldots i)^+ \to 1^+,2^+,\ldots,i^+\,] \; T_{n+1-i}[(i+1\ldots n+1)^+ \to i+1^+,\ldots,n+1^+]\,
\right\}\, . 
\end{multline}
This is simply the recursion relation for $T_{n+1}[(12\ldots n+1)^+ \to 1^+,2^+,\ldots, n+1^+]$ given in \eqref{eq:fragonestep} multiplied by $\left({z_{1}}/{z_{1\ldots n+1}}\right)^2$.  Thus, combining \eqref{eq:fragonestep}, \eqref{eq:newfrag} and \eqref{eq:newfragrecur} we get
\be
T_{n+1}[(12\ldots n+1)^- \to 1^-,2^+,\ldots, n+1^+] = \left(\frac{z_{1} }{z_{1\ldots n+1}}\right)^2 T_{n+1}[(12\ldots n+1)^+ \to 1^+,2^+,\ldots, n+1^+],
\ee
which proves \eqref{eq:newfrag} to be correct.

%%%%%%%%%%%%%%%%
\section*{Appendix B}
In this appendix we provide the proof for \eqref{eq:relation} which is done using induction. We begin by rewriting the \eqref{eq:relation} as
\be
\sum_{j=1}^k  \frac{z_{j+1\ldots k+1}^2}{z_{1\ldots j}}v_{j+1\;j}
 - \sum_{j=2}^{k} \frac{z_{1\ldots k+1}^2}{z_{1\ldots j+1}}v_{j+1\;j} \frac{v_{(1\ldots j) 1}^2}{v_{j+1(1\ldots j+1)}v_{j(1\ldots j)}} 
 -  z_{1\ldots k+1}  \frac{v_{(1\ldots k+1)1}^2}{v_{k+1(1\ldots k+1)}} =0.
 \label{eq:newRel1}
\ee
To get this we have substituted \eqref{eq:Mdef} into \eqref{eq:relation} and taken out some overall factors. We will label the left hand side of this equation as $f_k$ and assume that $f_k = 0$ for $k<k^\prime$, where $k^\prime$ is an arbitrary upper limit. If, under our assumption, $f_{k^\prime}=0$ for $k^\prime = k+1$ then it will be true that $f_k = 0$ for all $k$ subject to $f_{1} = 0 = f_2$.

Our current expression for $f_k$ is, however, too cumbersome to work with.  Thus, we will derive a simpler form.  We can combine the first and third terms by using \eqref{eq:rel1}.  Then using $v_{(1\ldots k+1) 1} = -v_{k+1(1\ldots k+1)}+v_{(k+1)1}$ and \eqref{eq:rel2} we get, after some manipulation,
\begin{align}
\sum_{j=1}^k  \frac{z_{j+1\ldots k+1}^2}{z_{1\ldots j}}v_{j+1\;j}
 -  z_{1\ldots k+1}  \frac{v_{(1\ldots k+1)1}^2}{v_{k+1(1\ldots k+1)}}
& =\sum_{j=1}^k  z_{j+1\ldots k+1} v_{j+1\;j} \left( \frac{z_{j+1\ldots k+1}}{z_{1\ldots j}} - \frac{v_{(1\ldots k+1)1}}{v_{k+1(1\ldots k+1)}} \right) \nonumber \\
& =-\sum_{j=2}^{k}\sum_{l=1}^{j-1} \frac{z_{j+1\ldots k+1} z_{l+1\ldots j}}{z_{1\ldots j} } \frac{v_{j+1\;j} \;v_{l+1\;l}}{v_{k+1(1\ldots k+1)}} 
+ \sum_{j=1}^{k-1}\sum_{l=j+1}^{k} \frac{z_{j+1\ldots k+1} z_{j+1\ldots l}}{z_{1\ldots j} } \frac{v_{j+1\;j}\; v_{l+1\;l}}{v_{k+1(1\ldots k+1)}}.
\label{eq:1&3_1}
\end{align}
Here we have changed the lower and upper limits of $j$ appropriately.  We now choose to change the order in which the sums are performed in the second term of \eqref{eq:1&3_1}.  I.e., we replace $ \sum_{j=1}^{k-1}\sum_{l=j+1}^{k} $ by $ \sum_{l=2}^{k}\sum_{j=1}^{l-1}$.  Furthermore, since $j$ and $l$ are dummy indices we can exchange them so that we can now combine the two terms in \eqref{eq:1&3_1} into a single double sum.  Thus, for the sum of the first and third terms of \eqref{eq:newRel1} we end up with
\begin{align}
-\sum_{j=2}^{k}\sum_{l=1}^{j-1} \frac{z_{j+1\ldots k+1} z_{l+1\ldots j}}{z_{1\ldots j} } \frac{v_{j+1\;j} \;v_{l+1\;l}}{v_{k+1(1\ldots k+1)}} 
+ \sum_{j=1}^{k-1}\sum_{l=j+1}^{k} \frac{z_{j+1\ldots k+1} z_{j+1\ldots l}}{z_{1\ldots j} } \frac{v_{j+1\;j}\; v_{l+1\;l}}{v_{k+1(1\ldots k+1)}}
= \sum_{j=2}^{k}\sum_{l=1}^{j-1}  \frac{z_{l+1\ldots j}^2z_{1\ldots k+1}}{z_{1\ldots l} \; z_{1\ldots j}} \frac{v_{j+1\;j} \;v_{l+1\;l}}{v_{k+1(1\ldots k+1)}} .
\label{eq:1&3_2}
\end{align}
We can also rewrite the second term in \eqref{eq:newRel1} in the following way,
\be
  -\sum_{j=2}^{k} \frac{z_{1\ldots k+1}^2}{z_{1\ldots j+1}}v_{j+1\;j} \frac{v_{(1\ldots j) 1}^2}{v_{j+1(1\ldots j+1)}v_{j(1\ldots j)}} =   -\sum_{j=2}^{k}\sum_{l=1}^{j-1} \frac{z_{1\ldots k+1}^2z_{l+1\ldots j}}{z_{1\ldots j+1}z_{1\ldots j}}v_{j+1\;j}v_{l+1\;l} \frac{v_{(1\ldots j) 1}}{v_{j+1(1\ldots j+1)}v_{j(1\ldots j)}} 
  \label{eq:2}
\ee
Finally, $f_k$ will be given by the sum of \eqref{eq:1&3_2} and \eqref{eq:2}.  Let us now define
\be
g_k \equiv \frac{v_{k+1(1\ldots k+1)}}{z_{1\ldots k+1}} f_k.
\ee
For $k^\prime=k+1$ we would then have
\begin{align}
g_{k^\prime}&=\frac{v_{k+2(1\ldots k+2)}}{z_{1\ldots k+2}} f_{k+1} \\
&= \sum_{j=2}^{k+1}\sum_{l=1}^{j-1}  \frac{z_{l+1\ldots j}^2}{z_{1\ldots l} \; z_{1\ldots j}} v_{j+1\;j} \;v_{l+1\;l} 
- \sum_{j=2}^{k+1}\sum_{l=1}^{j-1} \frac{z_{1\ldots k+2} z_{l+1\ldots j}}{z_{1\ldots j+1}z_{1\ldots j}}v_{j+1\;j}v_{l+1\;l} \frac{v_{(1\ldots j) 1}v_{k+2(1\ldots k+2)}}{v_{j+1(1\ldots j+1)}v_{j(1\ldots j)}} \nonumber \\
&=\frac{v_{k+2\;k+1}}{z_{1\ldots k+1}}\left(\sum_{l=1}^{k}  \frac{z_{l+1\ldots k+1}^2}{z_{1\ldots l} }  \;v_{l+1\;l} 
-\sum_{j=2}^{k}\frac{z_{1\ldots k+1}^2}{z_{1\ldots j+1}}v_{j+1\;j} \frac{v_{(1\ldots j) 1}^2}{v_{j+1(1\ldots j+1)}v_{j(1\ldots j)}}
- z_{1\ldots k+1}\frac{v_{(1\ldots k+1) 1}^2}{v_{k+1(1\ldots k+1)}}\right) \nonumber \\
&\;\;\;\;+\sum_{j=2}^{k}\sum_{l=1}^{j-1}  \frac{z_{l+1\ldots j}^2}{z_{1\ldots l} \; z_{1\ldots j}} v_{j+1\;j} \;v_{l+1\;l} 
- \sum_{j=2}^{k}\sum_{l=1}^{j-1} \frac{z_{1\ldots k+1} z_{l+1\ldots j}}{z_{1\ldots j+1}z_{1\ldots j}}v_{j+1\;j}v_{l+1\;l} \frac{v_{(1\ldots j) 1}v_{k+1(1\ldots k+1)}}{v_{j+1(1\ldots j+1)}v_{j(1\ldots j)}} \nonumber \\
&= \frac{v_{k+2\;k+1}}{z_{1\ldots k+1}} f_k + g_k
\end{align}
To get the third line we have used \eqref{eq:rel1} and $z_{1\ldots k+2} v_{k+2(1\ldots k+2)} = z_{1\ldots k+1} v_{k+2(1\ldots k+1)} = z_{1\ldots k+1} (v_{k+2\;k+1} + v_{k+1(1\ldots k+1)})$.  Our assumption $f_k = 0$ implies $g_k = 0$, since $v_{k+1(1\ldots k+1)}$ is in general non-zero.  Hence,  this shows that $f_{k+1}\propto g_{k+1} = 0$.  It is fairly easy to show that \eqref{eq:newRel1} is true for $k=1,2$.  Thus, we have shown that \eqref{eq:newRel1} is true for all $k\ge1$.

%% References with bibTeX database:

\bibliographystyle{elsarticle-num}
\bibliography{<your-bib-database>}

%% Authors are advised to submit their bibtex database files. They are
%% requested to list a bibtex style file in the manuscript if they do
%% not want to use elsarticle-num.bst.

%% References without bibTeX database:

\end{document}